# Collisionless damping of electron waves in non-Maxwellian plasma[1]

V. N. Soshnikov [2]

Plasma Physics Dept.,
All-Russian Institute of Scientific and Technical Information
of the Russian Academy of Sciences
(VINITI, Usievitcha 20, 125315 Moscow, Russia)

**Abstract**

*In this paper we have criticized the so-called Landau damping theory. We have analyzed solutions of the standard dispersion equations for longitudinal (electric) and transversal (electromagnetic and electron) waves in half-infinite slab of the uniform collisionless plasmas with non-Maxwellian and Maxwellian-like electron energy distribution functions. One considered the most typical cases of both the δ–type distribution function (the stream of plasma with monochromatic electrons) and distribution functions, different from Maxwellian ones as with a surplus as well as with a shortage in the Maxwellian distribution function tail. It is shown that there are present for the considered cases both collisionless damping and also non-damping electron waves even in the case of non-Maxwellian distribution function.*

**Introduction**

As it is well known, dispersion equations of plasma waves include indefinitely (logarithmically) divergent integrals (IDI), what implies only one sense: it is caused by an information lack in original kinetic and Maxwell equations. Therefore for the choice of way of taking these integrals it ought to supply original physical conditions with some additional ones. Our analysis of the simplest one-dimensional problem for a half-infinite isotropic plasma slab has shown that such additional clear physical conditions for considered cases are both absence of the so-called kinematical waves and absence of fast backward waves [1,2]. These conditions are satisfied for IDI's taken in the principal value sense following to the primary supposition by Vlasov [3] but are not satisfied for the contour sense of IDI's proposed by Landau, the latter ones leading to the ubiquitously recognized now effect of collisionless Landau damping for Maxwellian plasma [4]. Nevertheless collisionless damping is possible also for IDI's in the principal value sense, but only for the electron velocity distribution function $f_0$ different from Maxwellian one [5]. In this connection it is interesting to investigate the possibilities of collisionless wave damping or wave building up in a plasma with some more realistic non-Maxwellian distribution function on the base of before proposed by us method of characteristic (effective) velocity value [6]. This method simplifies drastically evaluation of IDI's in dispersion equations, though at the price of some uncertainty in the value of effective velocity $v_{eff}$ in the end result.

In the following we have used standard dispersion equations

---

$$G_{p_1 p_2} = 1 + \frac{\omega_L^2}{p_2} \int \frac{\partial f_0/\partial v_x}{p_1 + v_x p_2} dv_x = 1 + \frac{\omega_L^2}{k} \int \frac{\partial f_0/\partial v_x}{\omega - k v_x} dv_x \simeq 1 - \frac{\omega_L^2}{\omega^2 - k^2 v_{eff}^2} = 0, \qquad (1)$$

where $p_1 = i\omega$, $p_2 = -ik$ are Laplace transform parameters [6] (for the forward wave $\mathrm{Re}\, k > 0$) for longitudinal waves, and

$$G_{p_1 p_2} = p_2^2 - \frac{p_1^2}{c^2} + \frac{\omega_L^2 p_1}{c^2} \int v_z \frac{\partial f_0}{\partial v_z} \frac{1}{p_1 + v_x p_2} dv_z dv_x = -k^2 + \frac{\omega^2}{c^2} + \frac{\omega_L^2 \omega}{c^2} \int v_z \frac{\partial f_0(v_z v_x)}{\partial v_z} \frac{dv_x dv_z}{\omega - k v_x} = 0 \qquad (2)$$

for transversal waves with asymptotical solutions of the type $\sim \exp(i\omega t - ikx)$, where $\omega$ is given frequency and in the general case complex values $k$ are defined by dispersion equations, where $x$ is direction of wave propagation in the half-infinite plasma slab; $v_{eff}$ is some effective value of velocity $v_x$ in the farther described expressions. In the following $E$ is either transversal electric field $E_z$ for the case of transversal waves or longitudinal field $E_x$ for the case of longitudinal waves; $\omega_L$ is Langmuir electron frequency; $f_0$ is electron velocity distribution function normalized to unity.

In the case of Maxwellian distribution function approximate solutions of dispersion equations are

$$k^2 \simeq \frac{\frac{\omega^2}{c^2}\left(1 - \frac{\omega_L^2}{\omega^2}\right)}{1 + \frac{v_{eff}^2}{c^2}\frac{\omega_L^2}{\omega^2}} \simeq \frac{\omega^2}{c^2}\left(1 - \frac{\omega_L^2}{\omega^2}\right) \qquad (3)$$

(high velocity electromagnetic waves at account for the smallness of the term $\sim v_{eff}^2 k^2$ in denominator of expression (2));

$$k^2 \simeq \frac{\omega^2}{v_{eff}^2}\left(1 + \frac{v_{eff}^2 \omega_L^2}{c^2 \omega^2}\right) \simeq \frac{\omega^2}{v_{eff}^2} \qquad (4)$$

(low velocity transverse waves at account for the negligible smallness of the term $\omega^2/c^2$ in equation (2));

$$k^2 = \frac{\omega^2}{v_{eff}^2}\left(1 - \frac{\omega_L^2}{\omega^2}\right) \qquad (5)$$

(longitudinal electron waves according to equation (1)).

In these expressions one supposes that $v_{eff}$ equals approximately the value $\sqrt{v_x^2}$ within a factor $\sim(1,5 - 2)$.

It was shown before [5] that at the case of Maxwellian-like distribution function with some exceeding over Maxwellian distribution in the distribution tail there has place high collisionless damping of transverse electron waves of the low velocity electron mode. In the following we consider the most characteristic variants of collisionless solutions (as constituent parts of the general solution for low-collision plasma) at non-Maxwellian distribution functions.

**Effective value method**

In the preceding papers [1],[2],[5],[6],[7] we used an evident principle at taking integrals in dispersion equations: for some function $\Phi(x)$, for example Maxwellian-like distribution or any other function $w(x)$ there can be found such a value $x_{eff}$ that

$$\overline{\Phi(x)} = \Phi(x_{eff}), \qquad (6)$$

where in general case $x_{eff} \neq \bar{x}$. In the more general case with presence some additional parameters, e.g. $\Phi = \Phi(x,k)$, Eq.6 will transform to

$$\overline{\Phi(x,k)} = \Phi[x_{eff}(k), k], \qquad (7)$$

where the averaging integral is taken in the principal value sense. Thus functions of the type

$$\frac{w(x)\Phi(x)}{\omega - kx} \quad \text{or} \quad \frac{w(x^2)\Phi(x^2)}{\omega^2 - k^2 x^2} \qquad (8)$$

can be equalized zero near $\omega^2 - k^2 x^2 = 0$ at integrating in the principal value sense.

Since $x_{eff}(k)$ is variant at different $k$, in the first linear approximation one can suppose either

$$x_{eff}(k) \simeq x_{eff}(k_0)\left(1 + \alpha \frac{k}{k_0}\right) \qquad (9)$$

or

$$x_{eff}(k^2) \simeq x_{eff}(k_0^2)\left(1 + \alpha \frac{k^2}{k_0^2}\right), \qquad (10)$$

where $k_0$ or $k_0^2$ are some given values at real values $x_{eff}(k_0)$ or $x_{eff}(k_0^2)$ according to condition (6). The factor $\alpha$ is determined by the type of functions $\Phi(x)$ and $w(x)$; for the case of Maxwellian distribution it is assumed very small or equals zero due to very sharp lowering down in exponential tail of $w(x) = f_0(x^2)$.

At as a whole monotonically falling down function $f_0 \simeq \Phi(x^2)$ some surplus of $\Phi(x^2)$ in the tail of distribution function $f_0(x^2)$ leads to lowering $x_{eff}$, that is $\alpha < 0$, and vice versa at lowering $\Phi(x^2) > 0$ with the distribution tail below Maxwellian distribution $f_0$ it is equivalent to $\alpha > 0$. In the following we consider forward plasma waves with $\operatorname{Re} k_0 > 0$.

An important moment is here that parameter $k$ can be in general complex value determined as solution of dispersion equation. At real value $x_{eff}$ it means that factorization coefficient $\eta$ at $x_{eff}(k_0)$ in Eq.10 must be the real linear combination of products of complex and them conjugated values $\alpha$ and $k$ with $\operatorname{Re}\eta \neq 0$, $\operatorname{Im}\eta = 0$ (possible constrained as $\operatorname{Re}\eta > a$, where $a$ is some constant value), namely $\eta = \text{Re}(1 + \alpha k/k_0)$.

One can find an equation for determination complex $\alpha$ with calculation of $x_{eff}$ (either $x_{eff}^2$) by the direct numerical calculation of the principal values of integrals in dispersion equation at two selected values of complex $k$.

In the simplest interpretation one could find out even if qualitatively the fact itself of possibility of damping or building up forward wave with complex $k$, $\text{Re}\,k > 0$, neglecting further imaginary part $\text{Im}\,\alpha$. It is assumed that even such simplification gives key to understanding features of transversal electron (low-velocity) wave in the case of Maxwellian distribution function $\exp(-\xi^2)$ with some surplus over the latter at $\xi^2 \sim 4$, as it was discovered before [5] and in other analogous cases.

The real and imaginary parts of $k$ (and correspondingly $\alpha$) are connected to a certain extent by the requirement of absence of backward longitudinal waves or backward transverse electromagnetic waves. To satisfy this requirements the root of dispersion equation of for backward wave $\text{Re}\,k < 0$ (at $\omega > 0$) must be real value, that is with violation of the symmetry relative to $\pm \text{Im}\,k$ in forward and backward waves [7].

**Qualitative interpretation of damping of transverse low-velocity electron waves in the case of Maxwellian-like distribution function**

Par example, according to Eq.4 and Eq.10 we can obtain as one of variants

$$k^2 \simeq \frac{\omega^2}{v_{eff}^2 \left(1 + \alpha \frac{k^2}{k_0^2}\right)}, \qquad (11)$$

where $k_0 \equiv k(\alpha = 0) = \omega/v_{eff}(k_0)$, and $v_{eff}(k_0) \equiv v_0$ can be taken $\sim (1.5 - 2)\sqrt{v_x^2}$ corresponding with $\alpha \neq 0$. Thus

$$\frac{k^2}{k_0^2} \approx \left(\frac{\omega^2}{k_0^2 v_0^2}\right) \frac{1}{1 + \alpha \frac{k^2}{k_0^2}}, \qquad (12)$$

and

$$\frac{k^2}{k_0^2} = \frac{-1 + \sqrt{1 + 4\alpha}}{2\alpha}. \qquad (13)$$

The root sign is selected corresponding to a limit $k^2/k_0^2 \to 1$ at $\alpha \to 0$.

An addition of some positive surplus over Maxwellian distribution tail displaces $v_{eff}$ to more great values leading to smaller values of $|\alpha|$. At different $\alpha$ in Eq.13 we obtain the following table of $k/k_0$:

*Table 1*

| $\alpha$ | 0 | $-1/4$ | $-1/2$ | $-1$ | $-2$ |
|---|---|---|---|---|---|
| $k/k_0$ | 1 | 1.41 | $1.12 - 0.455i$ | $0.866 - 0.5i$ | $0.690 - 0.478i$ |

These results are in qualitative agreement with the strong damping of the transverse low-velocity electron mode in [5] at surplus over Maxwellian distribution $\sim \exp(-\xi^2)$ at the range $\xi^2 \sim 3-4$. Additions in the more remote tail of distribution function at $\xi^2 \geq 5$ lead to some lower $|\alpha| \neq 0$ up to disappearing of damping. The damping is absent at $\alpha \geq -1.4$, that is at small surpluses or some "gap" in the Maxwellian distribution tail with $\alpha > 0$. It ought also to note that the range $\xi^2 \sim 3-5$ in the case of gas discharge strongly ionized plasma corresponds usually to energy excitation of atomic and molecular levels up to energies of the order 10 eV.

**Fast (electromagnetic) waves in Maxwellian-like plasmas**

According to the effective velocity method dispersion equation of the fast velocity electromagnetic wave is (after corresponding replacing $v_{eff}^2$ in Eq.3) again

$$k^2 \simeq \frac{\frac{\omega^2}{c^2}\left(1-\frac{\omega_L^2}{\omega^2}\right)}{1+\frac{\omega_L^2}{\omega^2}\frac{v_0^2}{c^2}\left(1+\alpha\frac{k^2}{k_0^2}\right)} \simeq \frac{\omega^2}{c^2}\left(1-\frac{\omega_L^2}{\omega^2}\right), \qquad (14)$$

$k_0$ is determined as $k_0 \equiv k(\alpha = 0)$, and $v_0^2$ ought to be taken near the value $\overline{v_x^2}$. Thus, $k^2/k_0^2$ is in this case equal to $\simeq 1$ independently of $\alpha$ and distribution function.

**Longitudinal waves in Maxwellian-like plasma**

Analogously to the foregoing, dispersion equation (10) transforms to

$$1-\frac{\omega_L^2\left(\frac{m}{k_BT}\right)v_0^2\left(1+\alpha\frac{k^2}{k_0^2}\right)}{\omega^2-k^2\left[v_0^2\left(1+\alpha\frac{k^2}{k_0^2}\right)\right]} = 0, \qquad (15)$$

$v_0 \equiv v_{eff}$, and approximately it was supposed here

$$z \equiv \frac{m}{k_BT}v_0^2 = 1, \quad v_{eff} = v_0\left(1+\alpha\frac{k}{k_0}\right), \qquad (16)$$

$$k_0^2 \equiv k^2(\alpha = 0) = \frac{\omega^2}{v_0^2}\left(1-\frac{\omega_L^2}{\omega^2}\right). \qquad (17)$$

After elementary transformations one obtains solution

$$\frac{k^2}{k_0^2} \simeq \frac{1}{2\alpha}\left[-\left(1+\frac{\alpha\omega_L^2}{\omega^2-\omega_L^2}\right)^2 + \sqrt{\left(1+\frac{\alpha\omega_L^2}{\omega^2-\omega_L^2}\right)^2+4\alpha}\right], \qquad (18)$$

where root sign is selected so that at $\alpha \to 0$ also $k^2 \to k_0^2$. This solution is analogous to the solution (13) at the case of transverse low velocity electron wave with the possibility of strong damping at negative values $\alpha$.

## Longitudinal waves in plasma with $\delta$-like electron velocity distribution function

For the more clearness of the latter transformations with $\delta$-function the latter can be presented as a trapezium with smoothed angles and the width approaching to zero and the height approaching to infinity at keeping constant the area $S = 1$ (see Fig.1).

Integration by parts in expression (1) with

$$f_0(v_x) = \delta(v_x - v_0) \qquad (19)$$

leads to

$$\frac{\omega_L^2}{k}\int\frac{\partial f_0/\partial v_x}{\omega - kv_x}dv_x = \frac{\omega_L^2}{k}\left[\left.\frac{f_0}{\omega - kv_x}\right|_{-\infty}^{\infty} - \int\frac{kf_0}{(\omega - kv_x)^2}dv_x\right] =$$

$$= -\frac{\omega_L^2}{(\omega - kv_0)^2}. \qquad (20)$$

Thus dispersion equation at $p_1 = i\omega$, $p_2 = -i\omega$

$$G_{p_1 p_2} = 1 - \frac{\omega_L^2}{(\omega - kv_0)^2} = 0 \qquad (21)$$

has two solutions with forward waves

$$k = \frac{\omega \pm \omega_L}{v_0}. \qquad (22)$$

At $v_0 < 0$ and $\omega > \omega_L$ there are no solutions in the form of forward waves.

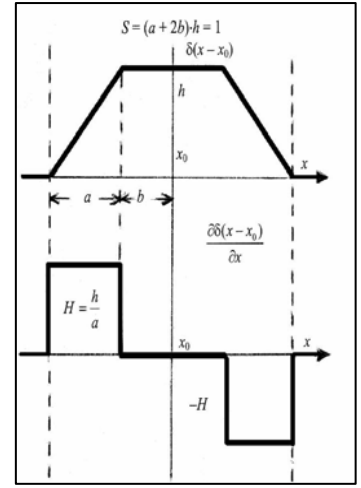

*Fig.1. Clear presentation of function $f_0 = \delta(x - x_0)$ as a trapezium with constant area $S = 1$; height, approaching to infinity; width, approaching correspondingly to zero. The derivative is also shown (not scaled). Integration by parts is possible because the function $\partial\delta(x - x_0)/\partial x \over c - x$ in all interval $(x \pm b)$ equals zero, it approaches also to zero after crossing over the point $x = c$ (the latter is an arbitrary constant) with its transition into region $x_0 \pm (a+b)$ at approaching to $\delta$-function limit.*

## Transverse waves in plasma with $\delta$-like electron distribution function

Since

$$f_0 = \delta(v_y)\delta(v_z)\delta(v_x - v_0), \qquad (23)$$

then

$$\int v_z \frac{\partial \delta(v_z)}{\partial v_z}dv_z = -1; \quad \int \delta(v_y)dv_y = 1, \qquad (24)$$

and dispersion equation is then written as

$$-\frac{c^2 k^2}{\omega^2} + 1 - \frac{\omega_L^2/\omega^2}{1 - \frac{k}{\omega}v_0} = 0. \qquad (25)$$

In the case of fast velocity (electromagnetic) wave with $kv_0/\omega \sim v_0/c \ll 1$ dispersion equation can be written approximately as

$$k^2 + \frac{\omega_L^2 v_0}{\omega c^2} k - \frac{\omega^2}{c^2}\left(1 - \frac{\omega_L^2}{\omega^2}\right) = 0 \tag{26}$$

with solution for the forward wave

$$k \simeq \frac{\omega}{c}\left(1 - \frac{\omega_L^2}{\omega^2}\right)^{1/2} - \frac{\omega_L^2 v_0}{2\omega c^2 \left(1 - \omega_L^2/\omega^2\right)^{1/2}}. \tag{27}$$

In the case of electron transverse wave, as before, neglecting in Eq.2 term $\omega^2/c^2$ (that is unit in Eq.25), supposing

$$k = k_0 + \tau, \quad |\tau| \ll k_0, \quad k_0 = \omega/v_0, \tag{28}$$

we obtain an equation for determining $\tau$

$$-\frac{\omega_L^2}{c^2} \frac{1}{1 - \frac{k_0 + \tau}{\omega} v_0} \frac{1}{(k_0 + \tau)^2} \simeq 1, \tag{29}$$

whence it follows

$$\tau \simeq \frac{v_0 \omega_L^2}{c^2 \omega} \ll k_0. \tag{30}$$

Thus, in transverse electron waves at $\delta$-like distribution function there is no exponential damping/growing of amplitudes.

**Longitudinal waves in Maxwellian plasma which moves with velocity $v_0$**

In this case distribution function at stream velocity $v_0$ (that is Maxwellian electron current in plasma medium) is

$$f_0 \propto \exp\left[-\frac{m}{2k_B T}(v_x - v_0)^2\right]. \tag{31}$$

At a more accurate consideration than expressed by relations (1) and in a more correspondence with the above-stated method of effective value, the analogue of expressions (1) will be

$$1 + \frac{\omega_L^2}{k} \int_{-\infty}^{+\infty} \frac{\partial f_0/\partial v_x}{\omega - k v_x} dv_x = 1 + \omega_L^2 \int_0^\infty \frac{u_x \partial f_0/\partial u_x}{(\omega - k v_0)^2 - k^2 u_x^2} du_x \approx 1 - \frac{\omega_L^2}{(\omega - k v_0)^2 - k^2 u_{xeff}^2} \cdot \frac{u_{xeff}^2}{u_0^2} = 0, \tag{32}$$

where

$$u_0^2 \equiv \frac{k_B T}{m}; \quad u_x \equiv v_x - v_0; \quad v_0/u_{xeff} \equiv \eta; \quad u_{xeff} > 0, \tag{33}$$

and $f_0$ is now renormalized to unity in integration limits $0 \div \infty$.

The solution of Eq.32 is now

$$k = -\frac{\omega}{u_{xeff}} \cdot \frac{\eta}{1-\eta^2} \pm \frac{\omega}{u_{xeff}} \cdot \frac{1}{1-\eta^2} \cdot \sqrt{\eta^2 + (1-\eta^2)\left(1 - \frac{\omega_L^2}{\omega^2} \cdot \frac{u_{xeff}^2}{u_0^2}\right)}, \quad (34)$$

where waves can exist only at $|\text{Re}\, k| < \infty$, that is $\eta^2 < 1$.

At $v_0 = 0$ there is

$$k_{v_0=0} = \pm \frac{\omega}{u_{xeff}} \cdot \left(1 - \frac{\omega_L^2}{\omega^2} \cdot \frac{u_{xeff}^2}{u_0^2}\right)^{1/2}. \quad (35)$$

This is generalization of Eq.5, where one assumed $u_{xeff}^2 \simeq u_0^2$. It ought to note that propagating waves can exist only at condition

$$\left(1 - \frac{\omega_L^2}{\omega^2} \cdot \frac{u_{xeff}^2}{u_0^2}\right) > 0. \quad (36)$$

One may assume that this relation is satisfied also at $v_0 \neq 0$. Then $k$ in Eq.34 is real value, that is the waves (if they exist) have no damping/growing features.

**Transverse electron waves in Maxwellian plasma which moves with velocity $v_0$**

Analogously to preceding section, with the distribution function

$$f_0 \propto e^{-\frac{m}{k_B T}(v_x - v_0)^2} \cdot e^{-\frac{m}{k_B T} z^2} \quad (37)$$

with following renormalization of $f_0$, we can rewrite expressions (2) in the form

$$-k^2 + \frac{\omega^2}{c^2} + \frac{\omega_L^2 \omega}{c^2} \int_{-\infty}^{+\infty} v_z \frac{\partial f_0(v_x, v_z)/\partial v_z}{\omega - k(u_x + v_0)} dv_x dv_z = -k^2 + \frac{\omega^2}{c^2} + \frac{\omega_L^2 \omega^2}{c^2} \int_0^\infty \frac{(\omega - kv_0) f_0(v_x)}{(\omega - kv_0)^2 - k^2 u_x^2} dv_x \approx$$

$$\approx -k^2 + \frac{\omega^2}{c^2} - \frac{\omega_L^2 \omega^2}{c^2} \cdot \frac{\omega - kv_0}{(\omega - kv_0)^2 - k^2 u_{xeff}^2} = 0; \quad u \equiv v_x - v_0, \quad (38a)$$

and then in equivalent form

$$-k^2\left[(\omega - kv_0)^2 - k^2 u_{xeff}^2\right] + \frac{\omega^2}{c^2}\left[(\omega - kv_0)^2 - k^2 u_{xeff}^2\right] - \frac{\omega_L^2 \omega^2}{c^2}(\omega - kv_0) = 0. \quad (38b)$$

For the low-velocity branch, neglecting terms of the order $\sim u_{xeff}^2/c^2$, $v_0^2/c^2$, one obtains dispersion equation in the form

$$(\omega - kv_0)^2 - k^2 u_{xeff}^2 = 0 \quad (39)$$

with its trivial solutions

$$k = \frac{\omega}{u_{xeff} + v_0}; \quad k = -\frac{\omega}{u_{eff} - v_0}. \tag{40}$$

Here $0 < u_{xeff} \ll c$, and slow-traveling waves exist at a condition $|\operatorname{Re} k| < \infty$ and also $\omega > \omega_L$ since the branch of the slow-velocity waves is coupled with the branch of the fast waves for which wave number is given by (3).

**Conclusion**

We have analyzed dispersion equations of electron waves in half-infinite slab of isotropic homogeneous plasma for characteristic variants of electron velocity distribution function: Maxwellian distribution with small (without sign changing of derivate) positive or negative additions in the distribution tail; monochromatic or Maxwellian-like beams moving as a whole with velocity $v_0$.

Exponential growing of traveling waves is observed in no variants, although this is not excluded in some other variants of distribution function.

Exponential damping appears only in the case of relatively small (without the change of derivative sign) positive addition in Maxwellian distribution tail $\exp(-\xi^2)$ at limited region $\xi^2 \simeq 3-4$ (see [5]) corresponding in the case of gas discharge plasma to electron energies from some eV up to $\sim 10$ eV.

In all cases we considered only forward waves in half-infinite slab of homogeneous electron plasma, what is physically justified for longitudinal electron and fast transverse electromagnetic waves. As it was shown in [8], slow backward waves which form at their reflection from the structure of fast waves can superpose over slow forward electron waves, therefore in the case of half infinite plasma slab slow transverse waves contain a constituent of oscillatory standing waves.

It appears however doubtless to be slow dependence of effective values $v_{eff}$, $u_{xeff}$ on $k$, so that

$$v_{eff}(k) \approx v_{eff}(k_0)\left(1 + \alpha \frac{k}{k_0}\right). \tag{41}$$

But this leads to more high orders of algebraic dispersion equations, that is to possible presence of some additional wave modes $k_n$ with a necessity of selection physical and unphysical roots.

With a series of our preceding papers this work is directed to the better understanding the nature of the real effect of collisionless damping of plasma waves together with some outlined ways of analytical and numerical solving corresponding them dispersion equations, including account for non-linear term and the case of low-collision plasma.